\documentclass[12pt]{article}
\begin{document}

\newcommand{\beq}{\begin{equation}}
\newcommand{\eeq}[1]{\label{#1}\end{equation}}
\newcommand{\bea}{\begin{eqnarray}}
\newcommand{\eea}[1]{\label{#1}\end{eqnarray}}
\newcommand{\Tr}{\mbox{Tr}\,}
\newcommand{\bm}[1]{\mbox{\boldmath{$#1$}}}

\begin{titlepage}

\begin{center}
{\Large \bf Causality Constrains Higher Curvature Corrections to Gravity}

\vspace{20pt}

{\large A. Gruzinov and  M. Kleban}

\vspace{12pt}

{ CCPP\\ 
Department of Physics\\ New York University\\
4 Washington Pl.\\ New York, NY 10003, USA }
\end{center}

\vspace{20pt}

\begin{abstract}
We show that causality constrains the sign of quartic Riemann corrections to the Einstein-Hilbert action. Our constraint constitutes a restriction on candidate theories of quantum gravity.

\end{abstract}

\end{titlepage}

\newpage
\section{Introduction}

It has been known for a long time that causality constrains effective low-energy Lagrangians (see \cite{aharonov}, or \cite{adams} for a recent discussion). For example, the Euler-Heisenberg correction to the Maxwell action is quartic in the field tensor $F$: $(FF)^2$ and $(F\tilde{F})^2$. Both terms must enter with the positive sign (as they actually do) to satisfy causality. Here by causality we simply mean that the group velocity of ultraviolet perturbations on arbitrary field backgrounds be less or equal to the speed of light. 
 
In type II string theory the first correction to the Einstein-Hilbert action is of  quartic order in the Riemann tensor $R$~ \cite{gross}, and in four dimensions takes the form $(RR)^2$ and $(R\tilde{R})^2$. We will show that both correction terms must enter with the positive sign (as they actually do) to satisfy causality.  Here by causality we again mean that the ultraviolet perturbations on an arbitrary background are light-like or subluminal on this background.

One should remember that causality violation and superluminal group velocity are not always equivalent \cite{aharonov}. Strictly speaking, our analysis does not say anything about causality. We merely show that positive quartic Riemann terms in the effective action give a negative (or zero) shift of the frequency of ultraviolet perturbations; if one of the quartic terms is negative, the frequency shift can be positive. The frequency shift is proportional to the cube of the wavenumber, so a positive frequency shift corresponds to a positive correction to the group velocity. 

We take positive frequency shifts as a proxy for causality violation. This makes sense, because the unmodified Einstein equation propagates the support of an infinitesimal perturbation on an arbitrary background exactly along the light cones of this background. In addition,  in the case of scalar and vector fields \cite{aharonov,adams} positive frequency shifts of this type do signal causality violation.

Terms in the effective action involving the Ricci tensor or scalar can always be removed by a redefinition of the metric \cite{gross}.  Therefore these terms can not be unambiguously determined from an $S$-matrix, for example from string theory, and we do not expect them to affect causality.  

In addition to quartic corrections we also discuss causality constraints for generic corrections of up to cubic order. Quadratic corrections to the action are topological in pure gravity in four dimensions and do not affect the dispersion relation (\S3 ), and so the leading correction is cubic in the Riemann tensor. We show that this correction does not affect the propagation velocity of ultraviolet perturbations. It remains to be seen, however, whether this is sufficient to ensure causality.

\section{Causality constraints on quartic Riemann}

Consider the Einstein-Hilbert action with quartic corrections:

\beq
S=\int d^4x\sqrt{-g}(R+c_1I_1^2+c_2I_2^2).
\eeq{qaction}
We use the mostly plus signature, $R$ is the Ricci scalar, and $I_a$ are curvature invariants:
\beq
I_1\equiv R_{\mu \nu \alpha \beta}R^{\mu \nu \alpha \beta}, ~~~~ I_2\equiv R_{\mu \nu \alpha \beta}\tilde{R}^{\mu \nu \alpha \beta},
\eeq{inv}
$R_{\mu \nu \alpha \beta}$ is the Riemann tensor, and $\tilde{R}_{\mu \nu \alpha \beta}$ is the dual tensor.  Equation (\ref{qaction}) is the most general quartic perturbation of the Einstein action because $I_1^2$ and $I_2^2$ are the only quartic invariants of the Riemann tensor in four dimensions (see for example \cite{Lifschitz}).  Varying the action gives the modified Einstein equation:
\beq
R_{\mu \nu}-{1\over 2}Rg_{\mu \nu}=8c_1\nabla ^\alpha \nabla ^\beta (I_1R_{\mu \beta \alpha \nu})+8c_2\nabla ^\alpha \nabla ^\beta (I_2\tilde{R}_{\mu \beta \alpha \nu})-{1\over 2}(c_1I_1^2+c_2I_2^2)g_{\mu \nu}
\eeq{einst}
where we have used the unperturbed solution $R_{\mu \nu} =0$ in writing the right hand side.

Let $g_{\mu \nu}$ be a solution of the modified Einstein equation (\ref{einst}).  An example is a plane wave background, for which $R_{\mu \nu}$ and all curvature invariants vanish.  We want to calculate the dispersion relation for the ultraviolet perturbations $\delta g_{\mu \nu}\equiv h_{\mu \nu}$ on this background. Here ultraviolet means that the wavelength of the perturbation is much smaller than the characteristic length of the background. We calculate the variation of the Einstein-Hilbert action to the leading (quadratic) order in the perturbation wavenumber $k_\mu$; the perturbation of the correction is also calculated to the leading (quartic) order in $k_\mu$. After we impose the transverse traceless conditions on the perturbation (again to leading order in $k_\mu$):
\beq
h=0, ~~~k_\mu h^{\mu \nu}=0,
\eeq{gauge}
we get 
\beq
k^2h_{\mu \nu}=64c_1S_{\mu \nu}S^{\alpha \beta}h_{\alpha \beta}+64c_2\tilde{S}_{\mu \nu}\tilde{S}^{\alpha \beta}h_{\alpha \beta},
\eeq{eigen}
where 
\beq
S_{\mu \nu}\equiv k^{\alpha}k^{\beta}R_{\mu \alpha \beta \nu}, ~~~\tilde{S}_{\mu \nu}\equiv k^{\alpha}k^{\beta}\tilde{R}_{\mu \alpha \beta \nu},
\eeq{S}
and we have used the gauge conditions (\ref{gauge}) and the unperturbed dispersion law $k^2=0$ in the right-hand side of (\ref{eigen}). 

The right-hand side of (\ref{eigen}) is of order $R^2k^4$, where $R$ is the characteristic curvature of the background. The terms of lower power in $k$ are not included. For example, perturbing $I_1^2g_{\mu \nu}$ gives the terms of order $R^3k^2$ and $R^4$ which we do not include in the ultraviolet eigenmode equation (\ref{eigen}).

The symmetric tensors $S_{\mu \nu}$, $\tilde{S}_{\mu \nu}$ are transverse and traceless:
\beq
S=0, ~~~k_\mu S^{\mu \nu}=0,~~~~~ \tilde{S}=0, ~~~k_\mu \tilde{S}^{\mu \nu}=0,
\eeq{Scond}
because the background has zero Ricci tensor, and the Riemann tensor is antisymmetric in the first and second pairs of indices. It follows that the gauge conditions (\ref{gauge}) agree with the eigenmode equation (\ref{eigen}). 

The eigenmode equation (\ref{eigen}) gives the ultraviolet dispersion law 
\beq
k^2=64c_1(S^{\alpha \beta}e_{\alpha \beta})^2+64c_2(\tilde{S}^{\alpha \beta}e_{\alpha \beta})^2,
\eeq{disp}
where $e_{\alpha \beta}$ is the space-like polarization tensor, $e_{\alpha \beta}e^{\alpha \beta}=1$ 

It is understood that the right-hand side is just a small correction to the standard dispersion law $k^2=0$, so that there are still just two frequency roots $k_0$ for a given spatial wavevector ${\bf k}$. Positivity of both $c_1$ and $c_2$ then ensures causality, in the sense that the dispersion law (\ref{disp}) gives only subluminal group velocities.

An explicit formula for the string theory quartic Riemann correction \cite{gross} can be found for example in \cite{giusto}. The correction is 
\bea
Y=t_8 t_8 R^4 &\ =& 12 (R_{\mu\nu\rho\sigma}R^{\mu\nu\rho\sigma})^2 + 
24 R_{\mu\nu\rho\sigma}R^{\rho\sigma\alpha\beta}R_{\alpha\beta\gamma\delta}R^{\gamma\delta\mu\nu}\nonumber\\
&-& 96 R_{\mu\nu\alpha\beta}{R^{\mu\nu}}_{\gamma\delta}R^{\rho\sigma\beta\gamma}{R_{\rho\sigma}}^{\delta\alpha}-
192 R_{\mu\nu\alpha\beta}R^{\mu\nu\alpha\gamma}R^{\rho\sigma\delta\beta}R_{\rho\sigma\delta\gamma}\nonumber\\
&+& 192 R_{\mu\nu\alpha\beta} R^{\rho\nu\gamma\beta} R_{\rho\sigma\gamma\delta}R^{\mu\sigma\alpha\delta}
+384 R_{\mu\nu\alpha\beta} R^{\rho\nu\gamma\beta}{R^{\sigma\mu}}_{\delta\gamma} {R_{\sigma\rho}}^{\delta\alpha}.
\eea{t8t8}
If we now compactify on a flat manifold to four dimensions this expression can be reduced to 
\beq
Y=24I_1^2+24I_2^2,
\eeq{Y}
corresponding to positive $c_1=c_2=24$, and to subluminal eigenmodes.  We have again kept only terms involving the Riemann tensor. Correction terms with powers of Ricci tensor are not uniquely determined and do not affect the propagation velocity of ultraviolet modes in pure gravity. 

\section{Cubic order}

The generic Lagrangian up to cubic order  contains fourteen terms \cite{metsaev}. At first order there is an Einstein-Hilbert term $R$; at second order there are three terms: $R^2$, $R_{\mu \nu}R^{\mu \nu}$, and $I_1$. The terms quadratic in Ricci scalar and tensor do not affect the vacuum Einstein equation $R_{\mu \nu}=0$, and the  $I_1$ term can also be dropped because the Gauss-Bonnet invariant $G_2\equiv I_1-4R_{\mu \nu}R^{\mu \nu}+R^2$ is the Euler invariant in 4 dimensions. 

There are ten cubic terms. We are interested only in the terms that are at most linear in the Ricci tensor and scalar. There are two such terms in 4 dimensions: $RI_1$ and $I_3$, where $I_3$ is the cubic curvature invariant:
\beq
I_3\equiv R_{\mu \nu \alpha \beta}R^{\alpha \beta}_{~~\rho \sigma}R^{\rho \sigma \mu \nu}.
\eeq{cinv}

First consider the cubic Riemann term
\beq
S=\int d^4x\sqrt{-g}(R+cI_3).
\eeq{caction}
The ultraviolet eigenmode equation is then 
\beq
k^2h_{\mu \nu}\propto S_{\alpha \mu}(k^2h^\alpha _\nu-k_\nu k^\beta h^\alpha _\beta )+S_{\alpha \nu}(k^2h^\alpha _\mu-k_\mu k^\beta h^\alpha _\beta ) .
\eeq{ceigen}
But now the gauge condition (\ref{gauge}) shows that the dispersion law remains unmodified: $k^2=0$. 

Similarly, the other invariant, $RI_1$,  does not modify the dispersion law. This does not prove causality, it just means that the leading-order geometrical optics cannot discover causality violations at cubic order.

\subsection*{Acknowledgments}
We thank Sergei Dubovsky, Gia Dvali, Giga Gabadadze, Alberto Nicolis, and Massimo Porrati for useful discussions. A.G. is supported by the David and Lucile Packard foundation.



\begin{thebibliography}{00}
\bibitem{aharonov}
  Y.~Aharonov, A.~Komar and L.~Susskind,
  ``Superluminal behavior, causality, and instability,''
  Phys.\ Rev.\  {\bf 182}, 1400 (1969).

\bibitem{adams}
  A.~Adams, N.~Arkani-Hamed, S.~Dubovsky, A.~Nicolis and R.~Rattazzi,
  ``Causality, analyticity and an IR obstruction to UV completion,''
  JHEP {\bf 0610}, 014 (2006)
  [arXiv:hep-th/0602178].

\bibitem{gross}
  D.~J.~Gross and E.~Witten,
  ``Superstring Modifications Of Einstein's Equations,''
  Nucl.\ Phys.\ B {\bf 277}, 1 (1986).



\bibitem{Lifschitz}
Lifshitz, E and Landau, L,
{\it The Classical Theory of Fields}, Butterworth-Heinemann (1980).

\bibitem{giusto}
  S.~Giusto and S.~D.~Mathur,
  ``Fuzzball geometries and higher derivative corrections for extremal
  holes,''
  Nucl.\ Phys.\ B {\bf 738}, 48 (2006)
  [arXiv:hep-th/0412133].


\bibitem{metsaev}
  R.~R.~Metsaev and A.~A.~Tseytlin,
  ``Curvature cubed terms in string theory effective actions,''
  Phys.\ Lett.\ B {\bf 185}, 52 (1987).










\end{thebibliography}
\end{document}